\documentclass[a4paper,12pt]{article}
\usepackage{graphicx,amsmath,amsfonts}
\def \ba{\begin{eqnarray}}\def\ea{\end{eqnarray}}
\def\bc{\begin{center}}\def\ec{\end{center}}
\def\nn{\nonumber\\}

\def\superscript#1{\raisebox{0.8ex}{#1}\hspace{0.05em}}

\title{\bf{Different approaches to calculate the $K^{\pm}\to \pi^{\pm}\pi^0e^+e^-$
decay width}}
\author{S. R. Gevorkyan, M. H. Misheva\superscript{\small }\thanks{On leave of absence from INRNE, BAS, Bulgaria}\\
 \small {Joint Institute for Nuclear Research, Dubna, Russia} }
\date{}
\begin{document}
\maketitle
\begin{abstract}
The rare $K^\pm\to\pi^\pm\pi^0 e^+e^-$ decay, currently under
analysis by the NA48/2 Collaboration, is considered. We have
performed two theoretical approaches to calculate the differential
decay width -- in the kaon rest frame, where we use
Cabibbo-Maksimovicz variables, and in the center-of-mass system of
the lepton pair. The latter essentially simplifies the computations.
A comparison between the two approaches has been performed. We have
also found the dependencies of the differential decay rate as a
function of the virtual photon and dipion system masses.
\end{abstract}
\section{Introduction}
For many years the radiative kaon decay $K^\pm\to\pi^\pm\pi^0\gamma$
has been considered as a good tool for studying the low energy
structure of QCD. The amplitude of this process consists of two
parts: a long distance contribution called inner Bremsstrahlung (IB)
and a direct emission (DE) part. IB contribution is associated with
the $K^\pm\to\pi^{\pm}\pi^0$ decay according to Low's
theorem~\cite{low58}, and DE can be calculated in the framework of
the Chiral Perturbation Theory (ChPT)~\cite{ecker88,ecker94}. In its
turn, the DE part is decomposed into electric and magnetic parts.
Despite the fact that the $K^\pm\to\pi^{\pm}\pi^0$ decay is
suppressed  by the $\Delta{I}=1/2 $ rule, the Bremsstrahlung
contribution is still
much larger than DE. \\
For the above mentioned radiative process DE is the region of relatively hard photons and
large angles between pion and photon. The
following variables are usually adopted~\cite{batley10}: the charged pion kinetic
energy in the kaon rest frame $T_c$; the  Lorentz invariant variable
$W^2=\frac{(p_1q)(p_Kq)}{m_{\pm}^2m_K^2}$, where $p_1,p_K,q$ are charged
pion, kaon and photon 4-momenta and $m_{\pm},m_K$ -- masses of $\pi^{\pm}$ and $K^{\pm}$ mesons.
These variables enable one to gain a distinction between DE and IB contributions
by means of the $W^2$-dependence of the decay width~\cite{christ67}.\\
As the DE piece is almost two order of magnitude smaller than the
Bremsstrahlung~\cite{PDG}, the correct consideration of
interference terms between IB and DE becomes crucial.
Recently, the NA48/2 Collaboration has measured the
interference of the Bremsstrahlung and electric parts for
$K^\pm\to\pi^{\pm}\pi^0\gamma$~\cite{batley10} and it has been shown that the
main contribution in DE comes from the magnetic part, which is more than
one order of magnitude larger than the electric contribution.\\
At present the NA48/2 Collaboration at CERN SPS is analyzing the
experimental data on the radiative decay with a virtual photon that
has not been observed up to now: \ba
K^\pm(P_K)\to\pi^\pm\pi^0\gamma^*(q)\to\pi^\pm(p_1)\pi^0(p_2)
e^+(k_1)e^-(k_2) . \label{mainpr} \ea The advantage of this decay in
comparison with the radiative decay with a real photon for the  DE
component extraction is obvious: the photon virtuality ($q^2$)
allows one to analyze the additional kinematical
\begin{figure}
\includegraphics[width=160mm]{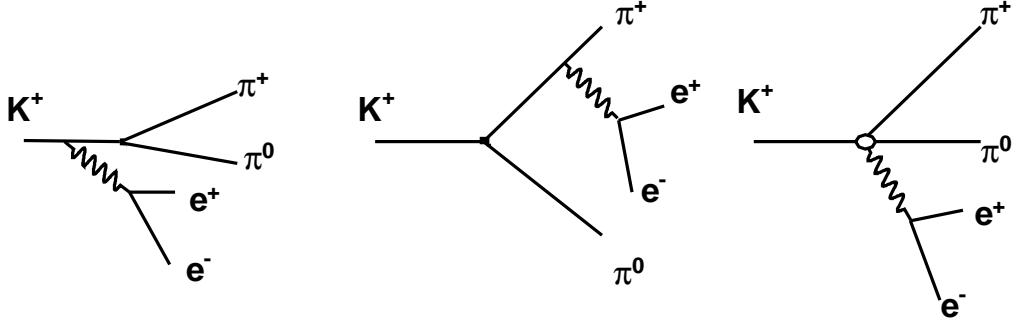}
\caption{\small \it{The first two diagrams represent the inner Bremsstrahlung
contribution. The third diagram corresponds to the direct emission.}}
\end{figure}
region which is absent in the case of real photons.\\
The solid theoretical base for this decay was developed in~\cite{pichl01},
where the DE contribution was calculated up to the next-to-leading
order (up to $O(p^4)$) in  ChPT.\\
The  essential step has been done recently~\cite{capp12}.
The authors have rewritten the matrix element and phase space in terms of
five independent variables relevant to the decay with a real photon and
have investigated the IB and DE contributions in different kinematical
regions.\\
Keeping  in mind the importance of the correct theoretical description of the decay
(1) and the necessity of taking into account all possible effects in view of
relative smallness of the DE contribution, we have slightly revised the
theoretical approach~\cite{capp12} to the $K^{\pm}\to \pi^{\pm}\pi^0e^+e^-$ decay, recalculating the
decay width by using the Cabibbo-Maksimovicz variables~\cite{cm65}.\\
As the next step, we have obtained the expression for the differential decay width
of $K^\pm\to\pi^\pm\pi^0 e^+e^-$ in the center-of-mass system (c.m.s) of the
lepton pair. Such approach simplifies the calculations and makes the exploration of the
rare process (\ref{mainpr}) more obvious.
Applying the obtained formulae, we calculate the contribution of
IB and DE to the differential decay rate as a function of the virtual
photon mass  $q^2$ and dipion mass $s_\pi$. As a result we have performed a comparison between
the two approaches in different frames of the $K^\pm\to\pi^\pm\pi^0 e^+e^-$ decay.
\section{Decay width}
The invariant amplitude of the decay (\ref{mainpr}) can
be parameterized as a product of leptonic and hadronic currents due to the covariance :
\ba
A=\frac{e}{q^2}j^{\mu}(k_1,k_2)J_{\mu}(p_1,p_2,q)
\ea
where  $p_1,p_2$ are the 4-momenta of charged and neutral pions,
$k_1,k_2$ -- the  leptons 4-momenta and $q = k_1 + k_2$ is the
momentum of the virtual photon. The leptonic current is:
\ba
j^{\mu}(k_1,k_2)=\bar{u}(k_2)\gamma^{\mu}v(k_1),
\ea
whereas the hadronic current is represented in terms of two electric form factors
$F_{1,2}$ and the magnetic one $F_3$:
\ba
J_{\mu}(p_1,p_2,q)=F_1p_{1\mu}+F_2p_{2\mu}+F_3\epsilon^{\mu\nu\alpha\beta}p_{1\nu}p_{2\mu}q_{\beta} .
\label{hadcurrent}
\ea
The decay width is given by the standard  expression:
\ba
d\Gamma=\frac{1}{2M_k}|A|^2 d \Phi .
\ea
The invariant phase space for the four-body decay is usually defined as:
\ba
d\Phi=(2\pi)^4\delta(P_k-p_1-p_2-k_1-k_2)\frac{d^3p_1}{2(2\pi)^3E_1}\frac{d^3p_2}{2(2\pi)^3E_2}
\frac{d^3k_1}{2(2\pi)^3\varepsilon_1}\frac{d^3k_2}{2(2\pi)^3\varepsilon_2} .
\label{phasespace}
\ea
The square of the leptonic current summed over spins is given as follows:
\ba
t^{\mu\nu}=\sum_{spins}{j^{\mu}j^{\nu}}=4(k_1^{\nu}k_2^{\mu}+
k_2^{\nu}k_1^{\mu}- g^{\mu\nu}(k_1k_2+m_e^2) )=
2(q_{\mu}q_{\nu}-k_{\mu}k_{\nu}-q^2g^{\mu\nu}),
\label{leptensor}
\ea
where  $k=k_1-k_2$ is the difference of leptons momenta, $m_e,m_{\pm},m_0$ are
electron, charged and neutral pion masses, correspondingly. \\
Introducing the relevant variables for the dipion  as $P=p_1+p_2$
and $Q=p_1-p_2$ and convoluting expression (\ref{leptensor}) with the square
of the hadronic current (\ref{hadcurrent}), we obtain the following expression for the
squared amplitude:
\ba
|A|^2& =&\frac{e^2}{q^4}\left(|A_E|^2+|A_M|^2+A_{EM}\right);\nn
|A_E|^2&=&
(-4m_{\pm}^2q^2+(qP+qQ)^2-(kP+kQ)^2)|F_1|^2 \nn
&+& (-4m_0^2q^2+(qP-qQ)^2-(kP-kQ)^2)|F_2|^2\nn
&+&(F_1F_2^*+F_1^*F_2)(-q^2(P^2-Q^2)+(qP)^2+(kQ)^2-(qQ)^2-(kP)^2);\nn
|A_M|^2&=&|F_3|^2
\{m_e^2[(16m_{\pm}^2m_0^2-(P^2-Q^2)^2)q^2-4m_{\pm}^2((qP)^2+(qQ)^2))\nn
&-&4m_0^2((qP)^2-(qQ)^2))+2(P^2-Q^2)((qP)^2-(qQ)^2)]\nn
&+&\frac{1}{4}(kP+kQ)^2((qp-qQ)^2-4m_0^2q^2))+\frac{1}{4}(kP-kQ)^2((qP+qQ)^2-4m_{\pm}^2q^2)\nn
&+&2((kP)^2-(kQ)^2)
(q^2P^2-q^2Q^2-(qP)^2+(qQ)^2)\};\nn
A_{EM}&=&\left((kP+kQ)(F_1^*F_3+F_1F_3^*)+(kP-kQ)(F_2^*F_3+F_2F_3^*)\right)
\epsilon^{\mu\nu\rho\sigma}k_{\mu}q_{\nu}P_{\rho}Q_{\sigma}.
\label{matrixkaoncms}
\ea
These formulae \footnote{There are several misprints in the
magnetic part of decay width Eq.(4.2) in
work~\cite{pichl01}). Moreover, the second term in the expression for
the electric form factor $F_2$ of this work Eq.(3.15) should be
reduced by twice to satisfy the Low's theorem.} are in accordance with expressions (19)
from~\cite{capp12}. The only difference is that we take into account the charged and
neutral pion mass difference in (\ref{matrixkaoncms}).\\
The electric form factors can be decomposed into Bremsstrahlung and direct emission
pieces: $F_i=F_i^B +F_i^{DE}$ while the magnetic form factor consists of
direct emission only $F_3=F_3^{DE}$. \\
Taking into consideration Low's theorem, the Bremsstrahlung part can be
written in terms of the matrix element for the  kaon decay into two
pions $M(K^+\to \pi^+\pi^0)$ and the sum of amplitudes corresponding to radiation of the
virtual photon by $K^{\pm}$--meson or charged pion:
\ba
M(K^+\to\pi^+\pi^0 \gamma^{\ast})_B= e M(K^+\to \pi^+\pi^0) \times
\left( - \frac{\epsilon_{\mu}P^{\mu}_k}{(P_k\cdot q) -
\frac{q^2}{2}} + \frac{\epsilon_{\mu}p^{\mu}_1} {(p_1 \cdot
q)+\frac{q^2}{2}}\right).
\ea
Comparing this expression with Eq.(\ref{hadcurrent}) for the hadronic current, one
immediately obtains  relations between the lowest order ~$O(p^2)$
contribution in electric form factors $F_1^B,F_2^B$ and decay
amplitude $M(K^+\to \pi^+\pi^0)$ ~\cite{capp12}:
 \ba
F_1^B&=&\frac{2ie(qP-qQ)}{(q^2+qQ+qP)(q^2+2qP)}M(K^+\to
\pi^+\pi^0),\nn
F_2^B&=&\frac{-2ie}{q^2+2qP}M(K^+\to \pi^+\pi^0).
 \ea
As it is mentioned above, the  matrix element of the $K^+\to\pi^+\pi^0$ decay and
the higher order terms ~$O(p^4)$ in the form factors caused by
direct emission can be calculated in ChPT ~\cite{pichl01,capp12}:
\ba
M(K^+\to\pi^+\pi^0)&=&\left(\frac{5}{3}G_{27}f_\pi(m_k^2-m_{\pm}^2)-f_\pi\delta{m^2}
(G_8+\frac{3G_{27}}{2})\right)e^{i\delta^2_0}\nn &=&|M(K^+\to
\pi^+\pi^0)|e^{i\delta^2_0} ,\nn
F_1^{DE}&=&-\frac{ieG_8e^{\delta^1_1}}{f_\pi}\left((qP-qQ)N_E^0+
\frac{4q^2N_E^1}{3}+4q^2L_9\right),\nn
F_2^{DE}&=&\frac{ieG_8
e^{\delta^1_1}}{f_\pi}\left((qP+qQ)N_E^0-\frac{2q^2N_E^{(2)}}{3}\right),\nn
F_3^{DE}&=&-\frac{2eG_8 e^{\delta^1_1}}{f_\pi}N_M^0,\nn
\delta{m^2}&=&m_{\pm}^2-m_0^2 .
\ea
Here $\delta^2_0$ and $\delta^1_1$ are the strong phases associated with
the interactions of the pions in the final state. In calculations we have used the values of
constants from Ref.\cite{capp12}.\\
These equations allow us to calculate the differential decay
width of the rare $K^{\pm}\to \pi^{\pm}\pi^0e^+e^-$ process \cite{pichl01}.\\
On the other hand, to describe the kaon decays to four particles in
the final state, it is enough to use five independent
variables as it was shown for the $K^{\pm}\rightarrow \pi^+ \pi^- e^{\pm}\nu$ ($K_{e4}$)
decay many years ago~\cite{cm65,pt68}.\\
Similarly to the $K_{e4}$ channel, we introduce five independent
variables which describe completely decay (\ref{mainpr}) -- dipion and
dilepton invariant masses, $s_{\pi}=(p_1+p_2)^2$ and $s_e=q^2=(k_1+k_2)^2$,
and three angles: $\theta_{\pi}$ -- the angle of the $\pi^{\pm}$ in the
($\pi^{\pm}\pi^0$) c.m.s with respect to the dipion flight direction; $\theta_e$ -- the
angle of the $e^+$ in the ($e^+e^-$) c.m.s with respect to the dilepton flight direction and
$\varphi$ -- the angle between dipion and dilepton planes. \\
Applying Lorentz transformations, one can express the covariant
scalar products in formulae (\ref{matrixkaoncms}) in terms of these variables:
 \ba
qP&=&\frac{m_k^2-s_{\pi}-s_e}{2}; \nn qQ&=& (qP) \cdot \frac{\delta
m^2}{s_{\pi}} + \frac{\beta_{\pi}
\lambda^{\frac{1}{2}}(s_{\pi},m_{\pm}^2,m_{K}^2)}{2}cos{\theta_{\pi}};\nn
kP&=&\frac{1}{2}
\beta_{e}\lambda^{\frac{1}{2}}(s_{\pi},m_K^2,q^2)cos{\theta_{e}};
\ea
\ba
kQ&=&\beta_e cos{\theta_e}[\frac{\delta m^2}{s_{\pi}}
\frac{\lambda^{\frac{1}{2}}(s_{\pi},m_K^2,q^2)}{2} +(qP)\beta_{\pi}
cos{\theta_{\pi}}]\nn &-& \beta_{\pi} \beta_e(q^2
s_{\pi})^{\frac{1}{2}} \sin{\theta_e} \sin{\theta_{\pi}}
\cos{\varphi}
\ea
\ba
\beta_{\pi}&=&\frac{\lambda^{\frac{1}{2}}(s_{\pi},m_{\pm}^2,m_0^2)}{s_{\pi}};\hspace{0.5cm}
\beta_e=\sqrt{1-\frac{4m_e^2}{s_e}};\nn
\lambda(x,y,z)&=&x^2+y^2+z^2-2xy-2xz-2yz .
\label{varkcms}
 \ea
Rewriting the invariant phase space (\ref{phasespace}) in terms of the variables introduced
above, we obtain the following:
 \ba
d\Phi &=&\frac{1}{4(4\pi)^6}\sqrt{1-\frac{4m_e^2}{s_e}}
\sqrt{1-\frac{(m_{\pm}+m_0)^2}{s_\pi}}\sqrt{1-\frac{(m_{\pm}-m_0)^2}{s_\pi}}
\sqrt{1-\frac{(\sqrt{s_\pi}+\sqrt{s_e})^2}{m_k^2}}\nn &\times&
\sqrt{1-\frac{(\sqrt{s_\pi}-\sqrt{s_e})^2}{m_k^2}}
ds_{\pi}ds_ed\cos{\theta_{\pi}}d\cos{\theta_e}d\varphi .
\label{phsp}
 \ea
Let us note that relations (\ref{varkcms}) and expression (\ref{phsp}) for the
phase space coincide with the relevant formulae in Ref.\cite{capp12}
if one rewrites them in terms of the corresponding
variables\footnote{Misprints in relations (15) of the
work~\cite{capp12} were corrected by the authors in Erratum.}.
\section{The decay width calculation in the dilepton center of mass system}
In the dilepton center of mass system ($\vec q=\vec k_1+\vec k_2=0$)
the virtual photon 4-momentum  $q=(\omega,0,0,0)$ and
$k=\omega(0,v\vec{n})$; $\omega$ is the virtual photon energy,
$\vec{n}$ is the unit vector and $v=\sqrt{1-\frac{4m_e^2}{\omega^2}}$
is the lepton velocity. The lepton tensor $t_{\mu \nu}$ in (\ref{leptensor})  has
the property $t_{00}=t_{0k}$ $(k=1,2,3)$ which essentially
simplifies the expression for the product of the lepton and hadron currents:
\ba
J_{\mu}J_{\nu}t_{\mu \nu}=s_e \left(|\vec{J}|^2 - (\vec{J}\vec{v})^2
\right)= s_e\left (|\vec{J}|^2 - v^2(\vec{J}\vec{n})^2 \right) .
\ea
The square of the matrix element reads:
\ba
\sum_{\lambda} |A|^2=\frac{2e^2}{s_e} \left(
|\vec{J}|^2 - \frac{(\vec{J}\vec{q})(\vec{J}\vec{q})}{s_e} \right).
\ea
Integrating this expression over the solid angle from the phase
space (\ref{phsp}), one obtains:
\ba
\sum_{\lambda} \int |A|^2 d\Omega_q=\frac{2\pi e^2}{s_e} |\vec{J}|^2
\left( 1-\frac{k^2}{3q^2}\right)=\frac{8\pi
e^2}{s_e}|\vec{J}|^2\left( 1-\frac{v^2}{3}\right).
\ea
The square of the hadron current is a function of pions
three-dimensional momenta $\vec{p_1},\vec{p_2}$ in the dilepton c.m.s:
\ba
|\vec{J}|^2=\vec{p_1}^{2}|F_1|^2+\vec{p_2}^{2}|F_2|^2+2(\vec{p_1}\vec{p_2})ReF_1F_2^{\ast}+
s_e(\vec{p_1}^2\vec{p_2}^2-(\vec{p_1}\vec{p_2})^2)|F_3|^2 .
\ea
To proceed further, we divide the pions momenta into longitudinal and
transverse parts. Using the Lorentz transformations, we express
them in terms of the pion momentum $p^{\ast}$ in the  dipion c.m.s:
\ba
p_{1L}=\gamma p^{\ast} \cos\theta+\beta E_1^{\ast},\nn p_{2L}=\gamma
p^{\ast} \cos\theta+\beta E_2^{\ast}, \nn
|\vec{p_{1\perp}}|=|\vec{p_{2\perp}}|=p^{\ast}\sin \theta .
\ea
where $\theta$ is the angle between the charged pion in the dipion c.m.s and
the dipion flight direction, $\gamma=\frac{M^2_K-s_{\pi}-s_e}{2\sqrt{s_{\pi}s_e}}$ is the
relevant Lorentz factor and $\beta=\sqrt{\gamma^2-1}$.
Now gathering the appropriate expressions, we obtain:
\ba
d \Gamma &=&\frac{{\alpha}^2}{4(4\pi)^3 M_K s_e} ( |F_1|^2
{\vec{p_1}}^2 + |F_2|^2 {\vec{p_2}}^2 + 2 (\vec{p_1}\cdot \vec{p_2})
Re(F_1F^{\ast}_2)\nn &+& s_e [ {\vec{p_1}}^2 {\vec{p_2}}^2 -
(\vec{p_1}\cdot \vec{p_2})^2 ] |F_3|^2 ) (1-\frac{v^3}{3})d s_{\pi}
d s_e d cos \theta ;\nn
  F^B_1 &=& \frac{2 i (\gamma E^{\ast}_2 -
\beta p^{\ast} cos{\theta}} {(\gamma E^{\ast}_1 + \beta p^{\ast}cos
\theta + \omega/2)(M_K^2-s_{\pi})} |M(K^{\pm}\rightarrow
\pi^{\pm}\pi^0)|e^{i{\delta}^2_0} ;\nn
F^B_2 &=& \frac{2
i}{(M_K^2-s_{\pi})} |M(K^{\pm}\rightarrow
\pi^{\pm}\pi^0)|e^{i{\delta}^2_0} ;\nn
F^{DE}_1 &=& \frac{2 i
G_8}{f_{\pi}}e^{i{\delta}^1_1} \left( N^0_E \omega(\gamma E^{\ast}_2
- \beta p^{\ast} cos \theta) +\frac{2}{3} {\omega}^2 N^1_E + 2 q^2
L_9 \right) ;\nn
F^{DE}_2 &=& \frac{2 i G_8}{f_{\pi}}e^{i{\delta}^1_1}
\left( N^0_E \omega(\gamma E^{\ast}_1 + \beta p^{\ast} cos \theta)
+\frac{1}{3} {\omega}^2 N^2_E \right);\nn
F^{DE}_3 &=& \frac{2 e G_8}{f_{\pi}}e^{i{\delta}^1_1}N^0_M .
\label{matrixlcms}
\ea
These formulae allow us to calculate the differential decay width
of the rare process (\ref{mainpr}) using the minimum set of variables
($s_e,s_\pi,\theta$).
\section{Numerical calculations and comparison of different
approaches}
First of all we have calculated the contribution of inner Bremsstrahlung and the full decay width
in the frameworks of the both approaches above mentioned.
For these calculations we have used set of constants from Ref.\cite{capp12},
the full kaon width and branching ratio to the hadronic decay
$K^\pm\to \pi^\pm\pi^0$ from \cite{PDG}. The full width of the
$K^\pm\to\pi^\pm\pi^0 e^+e^-$ decay  is  $\Gamma^{full}=2.231\times 10^{-22}$ MeV, whereas the
Bremsstrahlung contribution gives $\Gamma^{IB}=2.181\times
10^{-22}$ MeV. The branching ratio of the decay under consideration
is BR$(K^+\to \pi^+\pi^0e^+e^-)=419.33\times 10^{-8}$. The calculations in two
different approaches, the present one and that from Ref.\cite{capp12}, give the same numbers.
The computations using dilepton c.m.s (\ref{matrixlcms}) have an obvious advantage
in comparison with the kaon rest system (\ref{varkcms}) -- we need only three integrations
\begin{figure}[h]
\begin{minipage}[h]{.45\linewidth}
\includegraphics[scale=0.4]{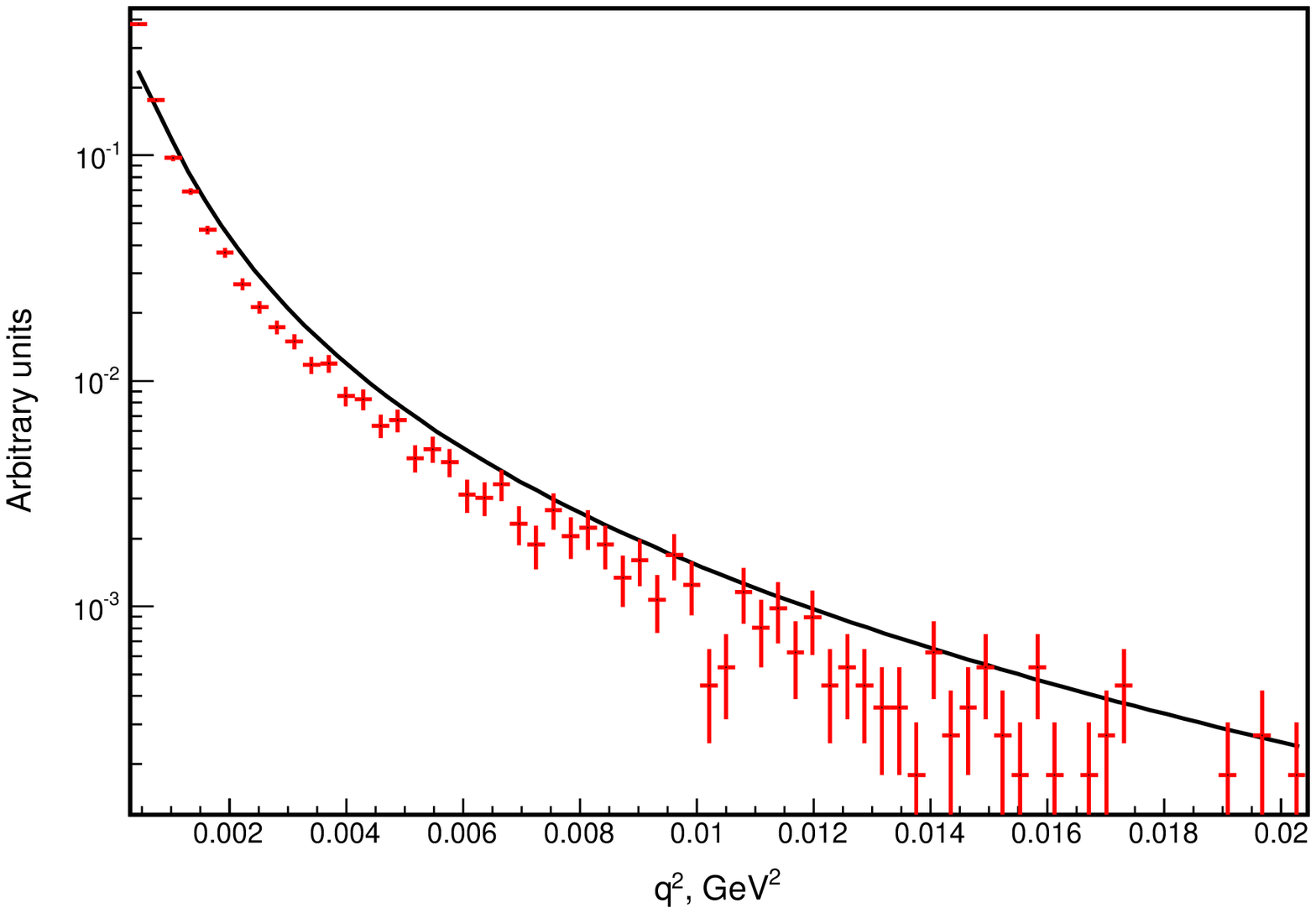}
\end{minipage}
\hspace{0.5cm}
\begin{minipage}[h]{.45\linewidth}
\includegraphics[scale=0.4]{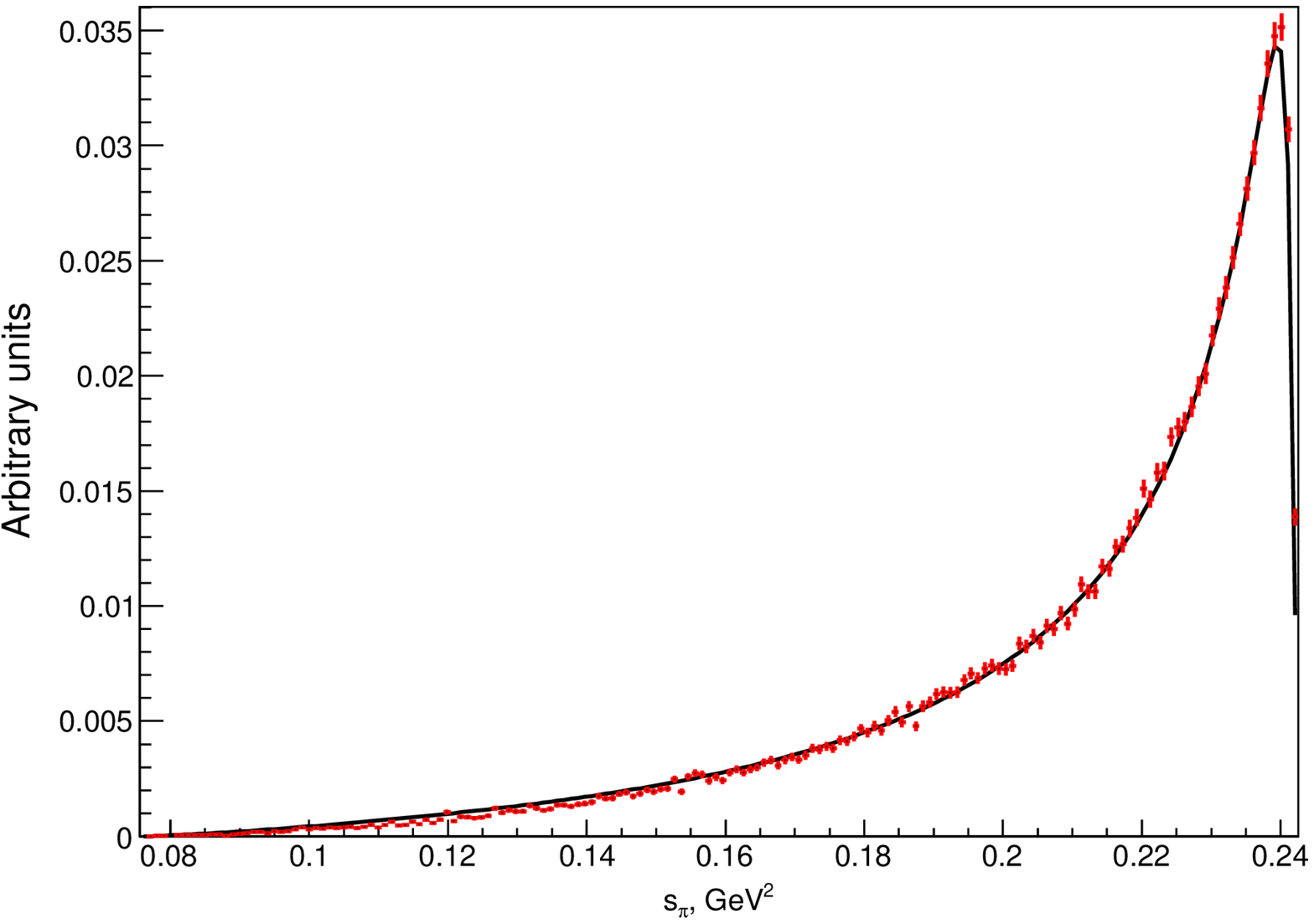}
\end{minipage}
 \caption{\small \it{Comparison of the differential decay width with respect to invariants
$q^2$ and $S_{\pi}$, obtained by theoretical calculation
in the lepton pair c.m.s (solid curve) and with the
MC generator in the kaon rest frame formulation (dots are
given with their statistical errors).}}
\label{compMClcms}
\end{figure}
for the full decay width calculation instead of the five integrals in Ref.\cite{capp12}.\\
The dependence of the decay width on invariant masses of dilepton
and dipion systems calculated by our formulae (\ref{matrixlcms}) has
been compared with the MC generator implemented by using the CERNLIB
library~\cite{CERN}, where the square of the matrix element from
(\ref{matrixkaoncms}) has been used. The result is shown in
Fig.\ref{compMClcms}.
As it is seen, the agreement is excellent.\\
Looking at the comparison between $\frac{d\Gamma^{IB}}{dq^2} (\frac{d\Gamma^{IB}}{ds_{\pi}})$
and $\frac{d\Gamma^{full}}{dq^2} (\frac{d\Gamma^{full}}{ds_{\pi}})$,
\begin{figure}[h]
\begin{minipage}[h]{.45\linewidth}
\includegraphics[scale=0.4]{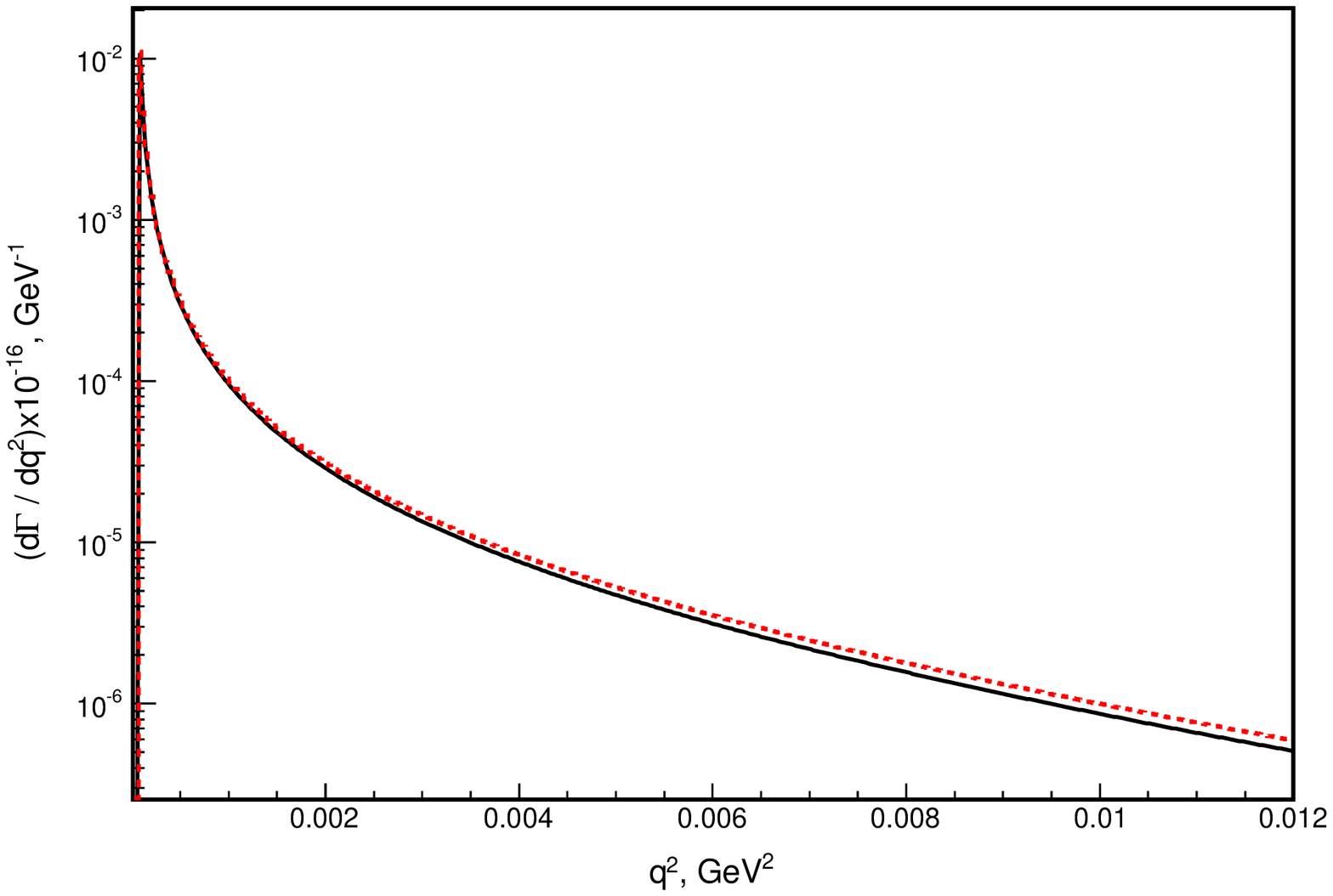}
\end{minipage}
\hspace{0.5cm}
\begin{minipage}[h]{.45\linewidth}
\includegraphics[scale=0.4]{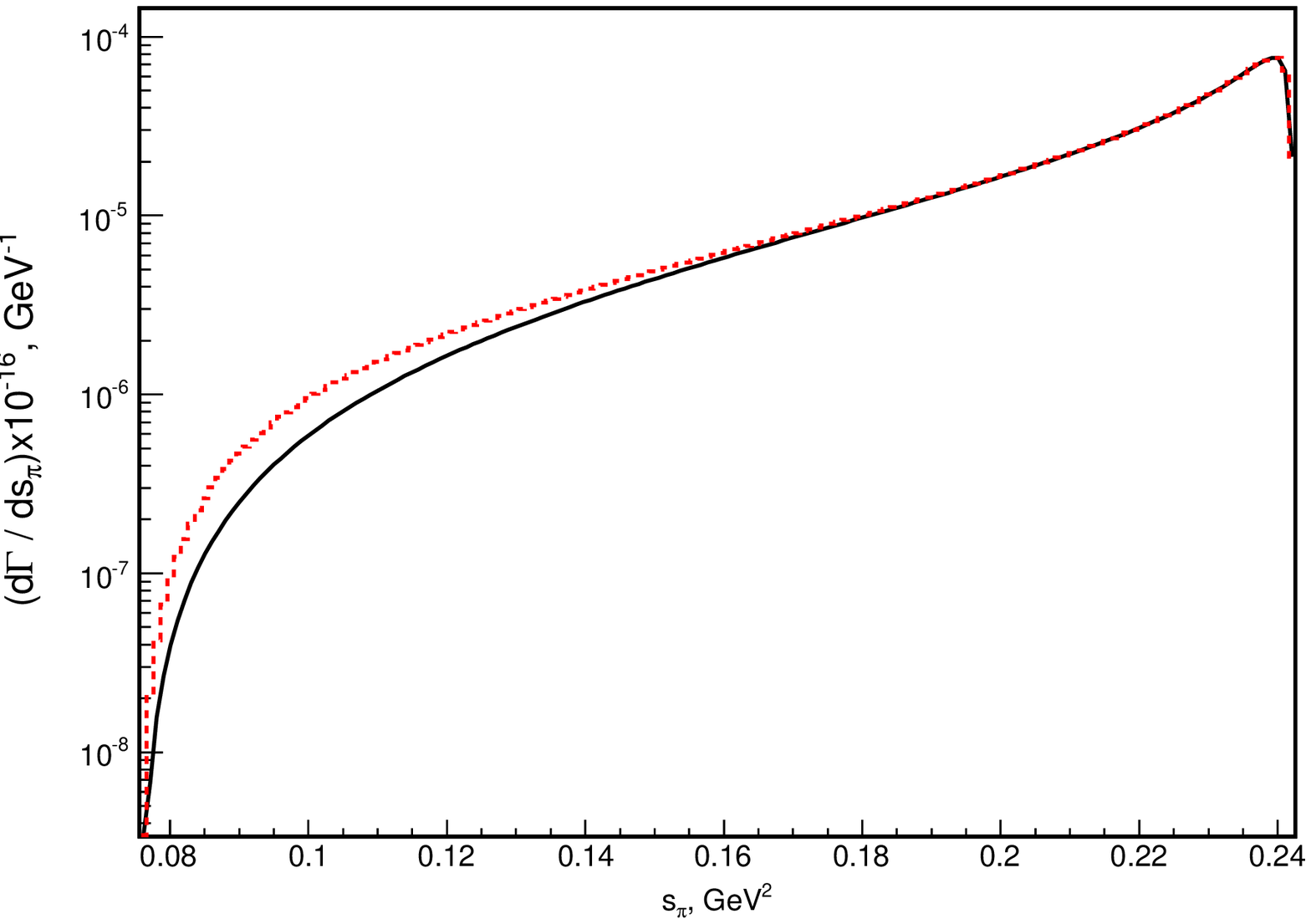}
\end{minipage}
 \caption{\small \it{Comparison between the decay width of IB contribution
(solid line) and the full decay width (dashed line)
with respect to the invariant masses of the dilepton and dipion systems.}}
\label{compibvstot}
\end{figure}
shown in Fig. \ref{compibvstot}, we see that the difference between
them due to the direct emission contribution is very small and it is
evident at large values of $q^2$ and in the region of small values
of $s_{\pi}$.

\section{Summary}
The general expression for the differential width of the $K^\pm\to
\pi^\pm\pi^0e^+e^-$ decay has been investigated in the kaon rest
frame and the dilepton c.m.s. Previously we have calculated the
differential decay width in terms of the Cabibbo-Maksymovicz
variables. We have also used the decay amplitude in the c.m.s of the
lepton pair, which is more convenient for computations. By means of
these expressions, we have calculated the branching ratio of the
$K^\pm\to\pi^\pm\pi^0 e^+e^-$ channel and obtained the dependencies
of the differential width on virtual photon mass $q^2$ and the
invariant mass of pion pair $s_\pi$ for inner Bremsstrahlung and
full decay widths. The comparison between the discussed approaches
is presented by using the dependence of the decay width on the
invariant masses $s_{\pi}$ and $q^2$.

\paragraph{Acknowledgements} We would like to thank O. Cata and M. Raggi
for very useful discussions.
We are also grateful to D. Madigozhin for his valuable consultations.
We appreciate the support of  V. Kekelidze and
Yu. Potrebenikov who sparked our interest to this work.


\begin{thebibliography}{99}
\bibitem{low58}F.E.Low,Phys.Rev.110,974 (1958)
\bibitem{ecker88}G.Ecker,A.Pich and E.De Rafael, Nucl.Phys.B303,665
(1988)
\bibitem{ecker94}G.Ecker,H. Neufeld and A.Pich, Nucl.Phys.B314,321
(1994)
\bibitem{batley10}J.R.Batley et al.Eur.Phys. J. C 68,75(2010)
\bibitem{christ67}N.Christ,Phys. Rev. 159,1292 (1967)
\bibitem{PDG}J. Beringer et al.(Particle Data Group),Phys. Rev.
D86,010001 (2012) and 2013 partial update for the 2014 edition.
\bibitem{pichl01}  H. Pichl, Eur.Phys.J.C 20,371 (2001)
\bibitem{capp12} L.Cappiello et al., Eur.Phys.J. C72, 1872
(2012); Erratum-ibid.C72,2208 (2012)
\bibitem{cm65}N. Cabibbo and A. Maksymowicz, Phys. Rev.137, 438 (1965)
\bibitem{pt68}A.Pais and S.B. Treiman,Phys.Rev. 168, 1858 (1968)
\bibitem{CERN}http://cernlib.web.cern.ch/cernlib
\end{thebibliography}
\end{document}